\documentstyle[11pt,aaspp4]{article}
\begin{document}
\slugcomment{To be published in the Astronomical Journal (April 2003)}

\title{A Survey of $z>5.7$ Quasars in the Sloan Digital Sky Survey II:
Discovery of Three Additional Quasars at $z>6$\altaffilmark{1}}

\author{Xiaohui Fan\altaffilmark{\ref{Arizona},\ref{IAS}},
Michael A. Strauss\altaffilmark{\ref{Princeton}},
Donald P. Schneider\altaffilmark{\ref{PSU}},
Robert H. Becker\altaffilmark{\ref{UCDavis},\ref{IGPP}},
Richard L. White\altaffilmark{\ref{STScI}},
Zoltan Haiman\altaffilmark{\ref{Princeton}},
Michael Gregg\altaffilmark{\ref{UCDavis},\ref{IGPP}},
Laura Pentericci\altaffilmark{\ref{Heidelberg}},
Eva K. Grebel\altaffilmark{\ref{Heidelberg}},
Vijay K. Narayanan\altaffilmark{\ref{Princeton}},
Yeong-Shang Loh\altaffilmark{\ref{Princeton}},
Gordon T. Richards\altaffilmark{\ref{Princeton}},
James E. Gunn\altaffilmark{\ref{Princeton}},
Robert H. Lupton\altaffilmark{\ref{Princeton}},
Gillian R. Knapp\altaffilmark{\ref{Princeton}},
\v{Z}eljko Ivezi\'{c}\altaffilmark{\ref{Princeton}},
W. N.  Brandt\altaffilmark{\ref{PSU}},
Matthew Collinge\altaffilmark{\ref{Princeton}},
Lei Hao\altaffilmark{\ref{Princeton}},
Daniel Harbeck\altaffilmark{\ref{Heidelberg}},
Francisco Prada\altaffilmark{\ref{Heidelberg},\ref{CalarAlto},\ref{LaPalma}}
Joop Schaye\altaffilmark{\ref{IAS}},
Iskra Strateva\altaffilmark{\ref{Princeton}},
Nadia Zakamska\altaffilmark{\ref{Princeton}},
Scott Anderson\altaffilmark{\ref{Washington}}.
Jon Brinkmann\altaffilmark{\ref{APO}},
Neta A. Bahcall\altaffilmark{\ref{Princeton}}
Don Q. Lamb\altaffilmark{\ref{Chicago}},
Sadanori Okamura\altaffilmark{\ref{UTokyo}},
Alex Szalay\altaffilmark{\ref{JHU}},
Donald G. York\altaffilmark{\ref{Chicago}}
}

\altaffiltext{1}{Based on observations obtained with the
Sloan Digital Sky Survey,
and with the Apache Point Observatory
3.5-meter telescope,
which is owned and operated by the Astrophysical Research Consortium;
on observations obtained at the W.M. Keck Observatory, which is operated as a scientific partnership
among the California Institute of Technology, the University of California and the National Aeronautics and
Space Administration, made possible by the generous financial support of the W.M. Keck
Foundation;
on observations obtained with the Hobby-Eberly Telescope, 
which is a joint project of the University of Texas at Austin,
the Pennsylvania State University, Stanford University,
Ludwig-Maximillians-Universit\"at M\"unchen, and Georg-August-Universit\"at
G\"ottingen;
and based on observations in the framework of the ``Calar Alto Key Project for SDSS Follow-up Observations'' (Grebel 2001)
obtained at the German-Spanish Astronomical Centre, Calar Alto Observatory, operated by the Max Planck Institute for
Astronomy, Heidelberg jointly with the Spanish National Commission for Astronomy.}

\newcounter{address}
\setcounter{address}{2}
\altaffiltext{\theaddress}{Steward Observatory, The University of Arizona,
Tucson, AZ 85721
\label{Arizona}}
\addtocounter{address}{1}
\altaffiltext{\theaddress}{Institute for Advanced Study, Olden Lane,
Princeton, NJ 08540
\label{IAS}}
\addtocounter{address}{1}
\altaffiltext{\theaddress}{Princeton University Observatory, Princeton,
NJ 08544
\label{Princeton}}
\addtocounter{address}{1}
\altaffiltext{\theaddress}{Department of Astronomy and Astrophysics,
The Pennsylvania State University,
University Park, PA 16802
\label{PSU}}
\addtocounter{address}{1}
\altaffiltext{\theaddress}{Physics Department, University of California, Davis,
CA 95616
\label{UCDavis}}
\addtocounter{address}{1}
\altaffiltext{\theaddress}{IGPP/Lawrence Livermore National Laboratory, Livermore,
CA 94550
\label{IGPP}}
\addtocounter{address}{1}
\altaffiltext{\theaddress}{Space Telescope Science Institute, Baltimore, MD 21218
\label{STScI}}
\addtocounter{address}{1}
\altaffiltext{\theaddress}{Max-Planck-Institut f\"{u}r Astronomie,
K\"{o}nigstuhl 17, D-69171 Heidelberg, Germany
\label{Heidelberg}}
\addtocounter{address}{1}
\altaffiltext{\theaddress}{Centro Astron\'omico Hispano-Alem\'an, Apdo 511, E-04080
Almer\'{\i}a, Spain
\label{CalarAlto}}
\addtocounter{address}{1}
\altaffiltext{\theaddress}{Current Address: Instituto de Astrof\'{\i}sica de Canarias,
E-38200 Tenerife and The Isaac Newton Group of Telescopes, Apdo 321,
E-38700 La Palma, Spain
\label{LaPalma}}
\altaffiltext{\theaddress}{University of Washington, Department of Astronomy,
Box 351580, Seattle, WA 98195
\label{Washington}}
\addtocounter{address}{1}
\altaffiltext{\theaddress}{Apache Point Observatory, P. O. Box 59,
Sunspot, NM 88349-0059
\label{APO}}
\addtocounter{address}{1}
\altaffiltext{\theaddress}{University of Chicago, Astronomy \& Astrophysics
Center, 5640 S. Ellis Ave., Chicago, IL 60637
\label{Chicago}}
\addtocounter{address}{1}
\altaffiltext{\theaddress}{Department of Astronomy and Research Center
  for the Early Universe, School of Science, University of Tokyo, Hongo,
  Bunkyo, Tokyo, 113-0033, Japan
\label{UTokyo}}
\addtocounter{address}{1}
\altaffiltext{\theaddress}{
Department of Physics and Astronomy, The Johns Hopkins University,
Baltimore, MD 21218, USA
\label{JHU}}

\begin{abstract}
We present the discovery of three new quasars
at $z>6$ in $\sim$ 1300 deg$^2$ of SDSS imaging data,
J114816.64+525150.3 ($z=6.43$), J104845.05+463718.3 ($z=6.23$)
and J163033.90+401209.6 ($z=6.05$).
The first two objects have weak Ly$\alpha$ emission lines;
their redshifts are determined from the positions of
the Lyman break. They are only accurate to $\sim 0.05$  and
could be affected by the presence of broad absorption line systems.
The last object has a Ly$\alpha$ strength more typical of lower redshift
quasars.
Based on a sample of six quasars at $z>5.7$ that cover 
2870 deg$^2$ presented in this paper and in Paper I, we estimate the
comoving density of luminous quasars at $z\sim 6$ and
$M_{1450} < -26.8$ to be
($8 \pm 3)  \times 10^{-10}$ Mpc$^{-3}$ (for $H_{0} = 50\ \rm km\ s^{-1} Mpc^{-1}$,
$\Omega = 1$).
HST imaging of two $z>5.7$ quasars and high-resolution ground-based
images (seeing $\sim 0.4''$) of three additional $z>5.7$ quasars 
show that none of them is gravitationally lensed.
The luminosity distribution of the high-redshfit quasar sample
suggests the bright end slope
of the quasar luminosity function at $z\sim 6$ is shallower than
$\Psi \propto  L^{-3.5}$ (2-$\sigma$),
consistent with the absence of strongly lensed objects.
\end{abstract}

\section{Introduction}

This paper is the second in a series presenting $i$-dropout
($z\gtrsim 5.7$) quasars selected from the multicolor imaging
data of the Sloan Digital Sky Survey (SDSS; \cite{York00},
\cite{EDR}).
In \cite{z58} and in Paper I (\cite{PaperI}), we presented the 
discovery of four quasars at $z=5.74$, 5.82, 5.99 and 6.28, respectively, 
selected from $\sim 1500$ deg$^2$ of SDSS imaging data in the
Northern Galactic Cap. 
The observations of these quasars provide important probes
to the high-redshift universe. 
They indicate the existence of supermassive black holes 
with $\rm M_{bh} \sim 10^{9-10} M_{\odot}$ at $z>6$ when 
the universe was less than 1 Gyr old.
The presence of strong metal emission lines in these quasars
suggests strong chemical enrichment and strong star-forming
activity in the quasar environment at early epoch.
The thickening of the Ly $\alpha$ forest, and the detection of
a complete Gunn-Peterson trough in the $z=6.28$ quasar
SDSS J1030+0524 (\cite{Becker01}, \cite{VLT}), indicate that the
ionizing background is declining rapidly, and the intergalactic
medium (IGM) neutral fraction is rising sharply at $z\sim 6$, suggesting
that we might be close to the epoch of reionization at this
redshift (e.g., \cite{Fan02}, \cite{CM2002}, 
\cite{Barkana2002}).

In this paper, we describe the discovery 
of three quasars at $z=6.05$, 6.23 and 6.43, selected
from $\sim 1300$ deg$^2$ of new SDSS imaging data.
The scientific objectives, photometric data reduction, target
selection and follow-up observation procedures are described
in detail in Paper I. The new observations used a slightly
relaxed set of selection criteria, and are outlined briefly
in \S 2.
We present the spectroscopic observations of the three new quasars,
and discuss their spectral properties in \S 3.
In \S 4, we combine the new quasars with those presented in Paper I
and recalculate the spatial density of luminous $z\sim 6$ quasars,
taking into account the effect of our color selection
criteria.
We also put constraints on the bright-end slope of the quasar luminosity
function based on this quasar sample.
At $M_{B} \sim -27$, the quasars presented in this paper and in Paper I
are the most luminous objects at high redshift, and
the probability of their being gravitationally lensed is boosted
by the magnification bias (\cite{TOG84}, \cite{WL2002},
\cite{CHS02}).
High-resolution images of these quasars are presented in \S 5
in order to uncover possible gravitational lensing.
None of the objects appears to be lensed under image resolution
from 0.1$''$ (HST) to 0.8$''$ (ground-based).
We use this observation to set a limit on the magnification bias
and thus to constrain the slope of the quasar luminosity function.
In a subsequent paper (Fan et al. 2003), we discuss the absorption
spectra of these quasars and their implication on the
reionization epoch in greater detail.

As in Paper I, the results are presented using two different
cosmologies: 
a $\Lambda$-dominated universe with $H_{0} = 65$ km s$^{-1}$ Mpc$^{-1}$,
$\Lambda = 0.65$ and $\Omega = 0.35$ (\cite{OS95}), which
is referred to  as the $\Lambda$-model; and
an Einstein-de Sitter universe with
$\Omega=1$ and $H_{0}$ = 50 km s$^{-1}$ Mpc$^{-1}$, which we refer to as
the $\Omega=1$ model in this paper.

\section{Candidate Selection and Identification}

The Sloan Digital Sky Survey is using
a dedicated 2.5m telescope and a large format CCD camera (\cite{Gunnetal})
at the Apache Point Observatory in New Mexico
to obtain images in five broad bands ($u$, $g$, $r$, $i$ and $z$,
centered at 3551, 4686, 6166, 7480 and 8932 \AA, respectively; \cite{F96})
in 10000 deg$^2$ of high Galactic latitude sky.
About $4000$ deg$^2$ of imaging data have been collected at the
time of this writing (November 2002).
The imaging data are processed by the astrometric pipeline
(\cite{Astrom}) and photometric pipeline (\cite{Photo}),
and are photometrically calibrated to a standard star network
(\cite{Smith02}, see also \cite{Hogg01}).

At $z\sim 5.7$, the Ly $\alpha$ emission line in the quasar spectrum
moves out of the $i$ band and into the $z$ band, the reddest filter in
the SDSS system. 
Quasars at higher redshift are characterized by their
very red $i-z$ color. 
They are very faint or are completely undetected in the $i$ band
due to the strong Ly $\alpha$ forest absorption in that band,
and become {\em $i$-dropout objects} with only one
measurable color (or its lower limit) in SDSS photometry.
As discussed in detail in Paper I, because of the
extreme rarity of very high-redshift quasars and the
overwhelming number of possible contaminants, mostly
cosmic rays in the $z$ band and very cool dwarfs (spectral
types L and T), the photometric selection procedure
of $z>5.7$ quasar candidates is quite complex.
The procedures include further image processing to eliminate 
single-band cosmic rays; independent $z$-band photometry
to improve the signal-to-noise ratio (S/N) of the $i-z$ color;
and $J$-band photometry, either from matching with the
Two Micron All Sky Survey (2MASS, \cite{2MASS}) catalog for bright sources, 
or independent $J$-band measurement for fainter sources, in order
to separate cool dwarfs from quasars, which have similar $i-z$
colors to those of $z>5.7$ quasars but much redder $z-J$ colors.

Paper I presents the results from a survey of $i$-dropout
candidates in $\sim 1500$ deg$^2$ of SDSS imaging carried  
out in the springs of 2000 and 2001.
In Spring  2002, we searched for $i$-dropout quasar
candidates in 48 new SDSS imaging runs.
These imaging data were taken between 27 April 2000 (Run 1411)
and 2 April 2002 (Run 3103).
We used the same criterion to decide which photometric runs
to include in the $i$-dropout survey as in Paper I: 
the $z$ band image quality, measured by 
the psfWidth parameter ($=1.06$ FWHM for
a Gaussian profile) in the fourth column of the SDSS camera,
should be  better than 1.8$''$. 
The median seeing in the $i$ and $z$ bands is $\sim 1.4''$
for the entire survey area.
The projection of the area covered by these runs is
illustrated in Figure 1.
This area overlaps with the
area covered by Paper I, mostly in the overlap
between adjacent SDSS strips and stripes (York et al. 2000).
Taking these overlaps into account, we find that the total
{\em new} area of the sky covered by these runs is
1320 deg$^2$, bringing the combined sky coverage of Paper I and
this paper to 2870 deg$^2$.

Figures 2 and 3 present the $i^*-z^*$ vs. $z^*$ color-magnitude
diagram
\footnote{Following Stoughton et al. (2002), we refer to the SDSS passbands as
$u$, $g$, $r$, $i$ and $z$. As the SDSS photometric calibration
system is still being finalized, the SDSS photometry presented here
is referred to as $u^*$, $g^*$, $r^*$, $i^*$ and $z^*$.}
and the $i^*-z^*$ vs. $z^*-J$ color-color diagram that
we use to select $i$-dropout candidates and to separate
high-redshift quasar candidates from cool dwarfs.
For quasars at $z>6.3$, a significant amount of $z$-band
flux is also absorbed by the Ly $\alpha$ forest absorption,
resulting in an increasingly red $z^*-J$ color towards
higher redshift (Figure 3). 
So in this Paper, we extend the $z^*-J$ color to be redder
than that in Paper I in order to search for quasars at
higher redshift.
The final color-selection criteria are:

\begin{equation}
 \begin{array}{l}
 (a)\ z^* < 20.2, \\
 (b)\ \sigma(z^*) < 0.1, \\
 (c)\ i^* - z^* > 2.2, \\
 (d)\ z^* - J < 1.5 + (i^* - z^* - 2.2) \times 0.35.
 \end{array}
\end{equation}

As in Paper I, we require that the object not be detected
in the $u$, $g$ or $r$ bands.
The photometric error in Eq. 1(b) refers to the error 
measured in the original SDSS imaging, not in the high S/N follow-up $z$-band
imaging.
In Eq. 1(d), the $z^*$ magnitude is in the  AB system while the
$J$ magnitude is in the Vega-based system.
We illustrate these color cuts in Figures 2 and 3.

A total of 98 $i$-dropout candidates were selected in the survey area.
The photometric and initial spectroscopic follow-up observations
were carried out over a number of nights between 26 December 2001 and 1 June 2002
using the ARC 3.5m telescope at the Apache Point Observatory.
Independent $z$ photometry was carried out using the
Seaver Prototype Imaging camera (SPICAM) in the SDSS $z$ filter on
the ARC 3.5m telescope.
Independent $J$ photometry was carried out using the GRIM II instrument
(the near infrared GRIsm spectrometer and IMager), also on the ARC 3.5m.
The initial spectroscopic follow-up observations were obtained
using the Double Imaging Spectrograph (DIS) on the ARC 3.5m.
For the three new quasars, further spectroscopic
observations were obtained using the Calar Alto 3.5m Telescope,
the Hobby-Eberly Telescope and the Keck II telescope
(see \S 3).
Descriptions of the instrument properties and data reduction
can be found in Paper I.

\section{Discovery of Three New Quasars at $z>6$}
 
Table 1 summarizes the classifications of the $i$-dropout
sample. Among the 98 $i$-dropout candidates,
20 are false $z$ band only detections which are most likely cosmic rays;
67 are M or L dwarfs (most of the M or L dwarfs are classified
photometrically based on their red $z^* - J$ colors). 
Among them, 
9 are likely T dwarfs (see
Knapp et al. 2003 for the infrared observations of
the new T dwarfs; several objects still lack 
proper infrared spectroscopy, so the T dwarf classification
is still preliminary). Two of
the candidates, SDSS J114845.05+463718.3 \footnote{The naming convention for SDSS
sources is SDSS JHHMMSS.SS$\pm$DDMMSS.S, and the positions are expressed in
J2000.0 coordinates. The astrometry is accurate to better than $0.1''$
in each coordinate.} (SDSS J1148+4637 for brevity), and
SDSS J104845.05+463718.3 (SDSS J1048+4637) are identified as
quasars at $z=6.43$ and 6.23, respectively.
These two objects are the only objects that satisfy the selection
criteria (Eq. 1) when improved $z$-band photometry is used.
In addition, we observed a number of objects with fainter
$z$-band magnitudes ($20.2 < z^* < 20.5$) and larger photometric
errors [$0.10 < \sigma(z^*) <0.12$] to test our ability to look
for faint $z\sim 6$ quasars using the SDSS photometry.
Among these faint candidates, one quasar (SDSS J163033.90+401209.6,
SDSS J1630+4012, $z=6.05$)
and one new T dwarf were discovered.
The survey at fainter magnitudes is not complete.

The finding charts and low resolution spectra of the three new
quasars are presented in Figures 4 and 5.
The spectra are flux-calibrated to match the observed $z$ band 
photometry.
Table 2 presents the photometric properties of the new quasars,
and Table 3 presents the measurements of their continuum properties.
Following Paper I, the quantity
$AB_{1280}$ is defined as the AB magnitude of the continuum 
at rest-frame 1280\AA, after correcting for interstellar
extinction using the map of \cite{Schlegel98}.
We extrapolate the continuum to rest-frame 1450\AA,
assuming a continuum shape $f_\nu \propto \nu^{-0.5}$ to 
calculate $AB_{1450}$.

\subsection{Notes on Individual Objects}

\noindent
{\bf SDSS J114816.64+525150.2 ($z=6.43$).}
This object is detected in two SDSS runs (Table 2) with
consistent photometry.
The discovery spectrum of SDSS J1148+4637 was obtained
using DIS on the ARC 3.5m on 21 April 2002.
The total exposure time was 3600 sec under partly cloudy skies.
This initial spectrum shows a strong, broad emission 
Ly $\alpha$+NV emission line at $\sim 9000$\AA\ and 
a clear Lyman break, indicating that it is at a redshift of $z\sim 6.4$,
thus making it the highest redshift quasar yet discovered.
Subsequently, we have obtained spectra with higher S/N using
three telescopes: 
On 29 April 2002, a moderate resolution ($R\sim 4000$) spectrum was
obtained with the Cassegrain Twin Spectrograph (TWIN) on
the Calar Alto 3.5m telescope. The total exposure time was
3.5 hours under good sky conditions.
The Calar Alto spectrum, which covers the spectral range of
8100\AA\ to 9800\AA, is shown in Figure 5. 
On 7 May 2002, a low resolution ($R\sim 500$) spectrum was
obtained with the Low Resolution Spectrograph (LRS, \cite{LRS}) on
the Hobby-Eberly Telescope (HET) with a total exposure time
of 5100 seconds.
On 8 May  and 10 May 2002, moderate resolution spectra ($R\sim 4000$) 
of SDSS J1148+5251 were obtained using the Echelle Spectrograph
Imager (ESI, \cite{ESI}) on the Keck II telescope. The sky was partly cloudy
with variable extinction of $\gtrsim 1$ mag.
The total exposure time was 3 hours, although the S/N of the
spectrum is worse than that in a 30 minute  exposure on Keck of a similarly
faint object under
good conditions, due to the clouds.
The final combined Keck spectrum (binned to 2\AA/pixel) is presented
in Figure 6.

While the presence of the Ly$\alpha$+NV emission line and Lyman
break is unambiguous, the determination of an accurate redshift
of this quasar is far from straightforward.
The Ly$\alpha$+NV emission line is quite weak.
The rest-frame equivalent width is $\sim 25$\AA, compared to
the average value of $\sim 70$\AA\ for quasars at $z\sim 4$
(Schneider, Schmidt, \& Gunn 1991, \cite{Fan01b}).
The blue side of the Ly $\alpha$ emission is almost totally absorbed
by the Ly $\alpha$ forest, and NV emission does not show a
separate peak.
No other emission line is detected
with the current S/N.
We adopt a redshift of $6.43 \pm 0.05$ using the peak
of Ly $\alpha$ emission and the onset of the Lyman break.
This value is quite uncertain, and a more accurate redshift can only
be determined using strong metal lines such as CIV, now redshifted well into
the near infrared $J$-band.
Note, however, high-ionization lines such as CIV are systematically 
blueshifted from the system redshift, which is better traced by low-ionization
lines such as MgII, by several hundred km s$^{-1}$ (e.g., \cite{Richards02}).
This will bias the redshift determined from CIV line at the $0.01 - 0.02$
level at $z\sim 6$.
We are  able to identify an intervening CIV doublet absorption system
at $z=4.95$ in both the Calar Alto and Keck spectra.
This absorption system is discussed further in a subsequent paper
(\cite{GP2}).

The fact that the Ly$\alpha$+NV emission line is very weak,
while not unique among high-redshift quasars
(\cite{BLLAC}), is intriguing
(see also the discussion of SDSS J1048+4637 below).
Clearly, a much larger sample is needed to study any trend in
the evolution of the emission line strength among the highest
redshift quasars.
The other possible explanation for the weak line 
is that the Ly$\alpha$ emission in SDSS J1148+5251
is affected by the presence of an intrinsic or associated NV absorption
system, as in the case of SDSS J1044--0125
(Fan et al. 2000), a BAL quasar at $z=5.74$
(\cite{Brandt01}, \cite{Djorgovski01}, \cite{Goodrich01}, \cite{Maiolino01}).
In this case, the redshift determined by the peak of what is left
of the Ly $\alpha$ emission line can be overestimated by $\sim 0.05$.
A $J$-band spectrum will be able to reveal the corresponding
CIV absorption, if this were indeed the case.

At $z^*=20.0$, $J=18.3$ and $M_{1450} = -27.6$ ($\Omega$-model, see Table 3),
SDSS J1148+5251 is extremely luminous.
Note that at this redshift, more than half of the flux in the $z$-band
is absorbed by the Ly$\alpha$ forest.
The continuum magnitude measured from the spectrum redward of Ly$\alpha$
emission ($AB_{1280} = 19.10$) is much brighter than
the broad-band $z$-band magnitude ($z^* = 20.0$) which is
affected by this absorption.
If not gravitationally lensed (see \S 5),
it is likely to be powered by a black hole of several billion solar masses,
within a dark matter halo of order $10^{13}$ $\rm M_{\odot}$ 
(see the discussion in Paper I). It is probably among the most luminous
and massive objects in the universe at $z>6$.
Detailed studies of its spectral properties, such as its X-ray and
sub-millimeter radiation, elemental abundances, and its environment,
will provide us important insights to early galaxy and star formation.

\noindent
{\bf SDSS J104845.05+463718.3 ($z=6.23$).}
The discovery spectrum of SDSS J1048+4637 was
obtained with DIS on the ARC 3.5m on 19 May 2002 with a total exposure time
of 3600 sec under partly cloudy skies.
Figure 5(b) shows a HET/LRS spectrum taken on 1 June 2002 with a total
exposure time of 3600 second.
This object also has a relatively weak Ly$\alpha$+NV emission line,
with a rest-frame equivalent width of $\sim 40$\AA, compared to an average
of $\sim 70$\AA\ at $z\sim 4$.
Using the peak of the Ly$\alpha$+NV emission and the onset of the Lyman break,
we estimate the redshift to be $z=6.23 \pm 0.05$.
Similar to the case of SDSS J1148+5251, this estimate is uncertain and
could be biased due to the presence of strong NV absorption.
We notice a possible absorption feature centered at 9750\AA.
It could be a SiIV BAL trough ($z_{\rm abs} = 5.95$)  if
confirmed by high S/N observations. 
At $J=18.4$ and $M_{1450} = -27.3$ (the $\Omega$-model), 
this is also a very luminous object.

\noindent
{\bf SDSS J163033.90+401209.6 ($z=6.05$).}
The discovery spectrum of SDSS J1630+4012 was obtained using
DIS on the ARC 3.5m on 2 June 2002 with a total exposure time of 1800 sec
under partly cloudy skies, resulting in very low S/N.
A higher S/N spectrum was obtained with HET/LRS on 3 July 2002 with
an exposure time of 3400 sec, and is shown in Figure 5(c).
Even with this moderate S/N and low resolution spectrum,  strong
Ly$\alpha$ and NV emission lines are clearly visible and the two peaks are
clearly separated.
The total rest-frame equivalent width of the Ly$\alpha$+NV lines
is $\sim 70$\AA, and the FWHM of the NV line is $\sim 4000\ \rm km\ s^{-1}$.
Using the NV peak, we estimate the redshift of SDSS J1630+4012 to be
$z=6.05 \pm 0.03$.
This object is the faintest quasar we have found so far at $z>5.7$.
Its $z$-band magnitude is 20.4, 0.2 mag fainter than the flux
limit of our complete survey.
It is only an 8-$\sigma$ detection in the SDSS $z$-band, and is very
close to its detection limit. It is encouraging that SDSS photometry 
still allows the detection and selection of $z\sim 6$ quasars at this
low S/N level. 
Studying quasars at lower luminosity requires deep imaging data such as
those provided by the SDSS Southern Survey with multiple
imaging (\cite{York00}).

Due to its proximity on the sky to Abell~2199, Abell~2197, and the $z=0.272$ 
active galaxy EXO~1627.3+4014, SDSS J1630+4012 serendipitously lies in 
several pointed observations made with the {\it ROSAT\/} Position Sensitive 
Proportional Counter (PSPC). We have analyzed PSPC sequences rp701507 and 
rp800644 which provide the most sensitive coverage of SDSS J1630+4012.
We do not find any highly significant detections in the standard 
{\it ROSAT\/} bands, although we note the presence of an $\approx 2\sigma$ 
source in the 0.5--2~keV band near SDSS J1630+4012 in rp701507; we shall 
conservatively set only an upper limit on the 0.5--2~keV flux. Applying the 
method of Kraft, Burrows, \& Nousek (1991), our 95\% confidence upper 
limit from rp701507 using a $70^{\prime\prime}$-radius aperture centered 
on SDSS J1630+4012 is 14.1 counts (the average effective exposure 
time in the aperture, accounting for vignetting, is 4440~s). We calculate 
an upper limit of $3.7\times 10^{-14}$~erg~cm$^{-2}$~s$^{-1}$ on the 
observed-frame, Galactic absorption-corrected, 0.5--2 keV flux (using a 
power-law model with a photon index of $\Gamma=2$ and the Galactic column 
density). Given the $AB_{1450}$ magnitude of SDSS J1630+4012, comparison 
with Figure~4 of Vignali et~al. (2003) shows that the X-ray upper limit
is consistent with X-ray observations of other $z>4$ quasars. 
SDSS J1630+4012 is not covered by the archival {\it Chandra\/} 
observations of Abell~2199, and we do not expect it to be covered by
the 2002 (still proprietary) {\it XMM-Newton\/} observation either.

From Figures 5 and 6, it is evident that the flux levels are consistent 
with zero (for the S/N in each case) in the wavelength range 
immediately blueward
of Ly$\alpha$ emission in SDSS J1048+4637 ($z=6.23$) and SDSS J1148+5251 ($z=6.43$).
Figure 6 shows clear presence of a complete Gunn-Peterson (1965) trough
from $\sim 8300$\AA\ to $\sim 8900$\AA\ in the spectrum
of SDSS J1148+5251, similar to the one detected in SDSS J1030+0524 
(Becker et al. 2001, Penterricci et al. 2002). 
While the S/N of the spectra is not as high for SDSS J1048+4637 and
SDSS J1630+4012, the average flux at $\lambda \sim 8200$\AA\ is
also at least a factor of 50 lower than the continuum level 
redward of Ly$\alpha$ emission.
In a subsequent paper (Fan et al. 2003), we discuss in detail the 
Gunn-Peterson effect along the different lines of sight and
its implication for the evolution of the ionizing background and
the epoch of reionization.

\section{Luminosity Function of $z\sim 6$ Quasars}

\subsection{Spatial Density of Luminous Quasars at $z\sim 6$}
In Paper I, we estimated the comoving density of quasars at $z\sim 6$
using a sample of four quasars.
We repeat this exercise here, adding the two additional quasars
SDSS J1148+5251 and SDSS J1048+4637 which have $z^* < 20.2$.
These six quasars form a complete sample 
satisfying  the selection criteria in Eq. (1) over
a total area of 2870 deg$^2$.
We do not use SDSS J1630+4012 as our follow-up observations
at $20.2 < z^* < 20.5$ are not complete.

As in Paper I, we calculate the selection function of
$z\sim 6$ quasars using a Monte-Carlo simulation of quasar
colors, taking into account the distribution of quasar emission line
and continuum properties, the Ly$\alpha$ absorption, the SDSS
photometric errors and the Galactic extinction. 
The selection function as a function of redshift $z$ and
absolute magnitude $M_{1450}$ is illustrated in Figure 7
for the $\Lambda$-model.
At $z>6.2$, the selection function changes somewhat from 
that in Figure 8 of Paper I:
the relaxed color cut in the $i^*-z^*$ vs. $z^*-J$ color-color
diagram results in a higher completeness at the highest redshift bins.
Note that in principle the selection allows discovery of quasars
at redshift as high as $z\sim 6.6$, but only with an extremely high luminosity
of $M_{1450} \lesssim -28$.

We derive the total
spatial density of quasars at $z\sim 6$ using the $1/V_{a}$ method:
\begin{equation}
V_a = \int_{\Delta z} p(M_{1450}, z) \frac{dV}{dz} dz,
\end{equation}
where $p(M_{1450}, z)$ is the selection function and the integral 
extends over the redshift range $5.7 < z < 6.6$.
The total spatial density in the sample is estimated by:
\begin{equation}
\rho = \sum_i \frac{1}{V_a^i}.
\end{equation}

 Using the selection function in Figure 7, we find  that at
the average redshift of $\langle z \rangle = 6.08$,
$\rho (M_{1450} < -26.8) = (8 \pm 3) \times 10^{-10}$ Mpc$^{-3}$
for the $\Omega=1$ model,
and $\rho (M_{1450} < -27.1) = (5  \pm 2) \times 10^{-10}$ Mpc$^{-3}$
for the $\Lambda$-model.
The results are consistent with those in Paper I with smaller
error bars.
In Figure 8, we show the evolution of the spatial density of luminous
quasars by combining the measurement here with the low redshift
measurement from the 2dF survey (\cite{2dF}) at $z<2.2$, 
from \cite{SSG} at $2.7 < z < 4.8$ and from Fan et al. (2001a) at
$3.6 < z < 5.0$.
The measurement at $z\sim 6$ is consistent with the decline of 
quasar density observed at $2 < z < 5$.
The comoving density of luminous quasars at $z\sim 6$ is 20 times
smaller than that at $z\sim 3$.

\subsection{Constraining the Slope of Quasar Luminosity Function}

The shape of the quasar luminosity function at high redshift provides
important constraints on the quasar contribution to the UV
ionizing background (e.g. \cite{MHR99}, Paper I) and on 
models of quasar formation and evolution (e.g. Haiman \& Loeb 1998).
Figure 7 shows that our color selection is sensitive to
quasars with $M_{1450} \lesssim -26.7$ at $z\sim 6$ ($p > 40\%$,
for the $\Lambda$-model).
However, none of the six $z>5.7$ quasars in our complete sample
has $M_{1450} > -27.1$.
How does this constrain the bright-end slope of the high-redshift
quasar luminosity function?

We can calculate constraints on the bright-end luminosity function, assuming
it is represented by a power-law: $\Psi \propto L^{\beta}$.
In Paper I, we placed a loose limit of $\beta > -3.9$ at the 2-$\sigma$ level
by looking at the luminosities of the two brighter vs. two fainter 
quasars in the original sample.
Here we try to directly fit the luminosity distribution of the
six quasars in the sample by a maximum likelihood estimate,
following Marshall (1985) and Fan et al. (2001b).
Given the limited redshift and luminosity range and the small 
number of objects in
the sample, we do not attempt to fit the redshift evolution and
assume the quasar luminosity function to be a single power law at
the bright end:
\begin{equation}
\Psi(M_{1450}) = \Psi^* 10^{-0.4[M_{1450}+26](\beta+1)}.
\end{equation}
The likelihood function can be written as: 
\begin{equation}
S = - 2 \sum_i^N \ln [\Psi(L_i, z_i) p(L_i, z_i)] +
2 \int\!\! \int \Psi(L, z) p(L, z) \frac{dV}{dz} dL\ dz.
\end{equation}

For the $\Omega=1$ model, we find 
$\Psi^* = (1.5^{+0.8}_{-0.5}) \times 10^{-9} \rm Mpc^{-3}$ 
(68\% confidence level) and for the $\Lambda$-model,
$\Psi^* = (1.3^{+0.7}_{-0.4}) \times 10^{-9} \rm Mpc^{-3}$.
The best-fit bright-end slope is $\beta = -2.3$,
with a 68\% confidence range of [--1.6, --3.1] and
a 95\% confidence range of [--1.3, --3.5].
The slope of $z\sim 6$ quasar luminosity function is 
consistent with that measured at $z\sim 4$ (Fan et al. 2001b,
$\beta = -2.6 \pm 0.2$), and the best-fit value is shallower than
the bright-end slope at $z<2.2$ from the 2dF survey, $\beta = -3.4$.
A slope steeper than $\beta = -3.5$ is excluded
at the 2-$\sigma$ level at $z \sim 6$ based on these calculations.

The constraint presented here is still very tentative, due 
both to the small number statistics (the sample only contains six objects)
and to the difficulty in estimating the selection function at
a magnitude close to the photometric detection limit.
The best-fit slope depends sensitively on the selection probability near the
detection limit.
We now attempt to constrain this slope independently using the magnification
bias effect of gravitational lensing.

\section{Gravitational Lensing of $z\sim 6$ Quasars}

The luminous quasars presented in this paper and in Paper I are
likely to be in black holes with masses up to several billion
solar masses. The presence of such massive black holes poses a challenge to 
models of structure formation and black hole formation at high
redshift. It is therefore important to ask whether any or all of
these quasars are magnified by gravitational lensing
(\cite{WL2002}, \cite{CHS02}).
At $M_{1450} \lesssim -27$, these quasars are located at the
bright end of the quasar luminosity function. 
Gott et al. (1984) first pointed out that in a fluxed-limited sample, the 
lensing probability of the brightest observed objects is boosted by
magnification bias.
Pei (1995) discussed the effect of magnification bias on a flux-limited
sample of high-redshift quasars.
In this section, we first present the HST and $K'$-band Keck imaging of 
our $z\sim 6$ quasars to search for multiply-imaged quasars
in our sample, and use the non-detection
of lensed quasars to put a constraint on
the amount of magnification bias and the slope of the quasar luminosity
function at $z\sim 6$.

\subsection{High-resolution Imaging of $z>5.7$ Quasars}

In HST Cycle 11, we carry out a snapshot survey of 
$\sim 200$ SDSS quasars at $z>4$ to search for multiply-imaged
quasars using the High Resolution Channel (HRC) of the
Advanced Camera for Surveys (ACS)
\footnote{Based on observations made with the NASA/ESA Hubble Space Telescope, obtained at the Space Telescope Science Institute, 
which is operated by the Association of 
Universities for Research in Astronomy, Inc., 
under NASA contract NAS 5-26555. 
These observations are associated with proposal 9472.}. 
Details of this snapshot survey are described in \cite{HST}.
At the time of this writing (Nov 2002), two of the $z>5.7$ quasars,
SDSS J0836+0054 ($z=5.82$) and SDSS J1030+0524 ($z=6.28$), have been
observed. The exposure time is 40 minutes for each quasar in the
$z'$-band, resulting in a limiting magnitude of $z' \sim 23$ at 10-$\sigma$
level.
At the HST resolution (0.1$''$), 
both quasars are consistent with being unresolved
point sources.

The Keck $K'$-band image of SDSS J1044--0125 ($z=5.74$) is described
in Fan et al. (2000). 
Keck images of the three other $z>5.7$ quasars, SDSS J1030+0524 ($z=6.28$),
SDSS J1048+4637 ($z=6.23$) and SDSS J1148+5251 ($z=6.43$),
were obtained in
two runs in May and June 2002, 
using the Near Infrared Camera (NIRC)
on the Keck~I telescope.
All observations were carried out under photometric skies and good
seeing, with total exposure time of 900 second for each objects.
The data were flattened,
sky-subtracted, shifted, and stacked using the DIMSUM package in IRAF.
The seeing of these observations is $0.40'' - 0.45''$.
The $K'$-band magnitudes for SDSS J1030+0524, SDSS J1048+4637, and SDSS J1148+5251,
are 17.64, 16.99 and 16.91, respectively.
The images show all quasars to be unresolved point sources.
We also look for close companions, and none was found with $K'>21$ 
within 10$''$ from the quasars.

For the other two known quasars at $z>5.7$, SDSS J1306+0356 ($z=5.99$) and
SDSS J1630+4012 ($z=6.05$), the best images we have are from the
ARC 3.5m under seeing of $0.7'' - 0.8''$.
On these images, the quasar also appear to be an unresolved point
sources.
Note that SDSS J1630+4012 is more than 30$'$ away from the center of
the two galaxy clusters Abell 2197 and Abell 2199  and is not likely 
to be lensed by them.

None of the $z>5.7$ quasars appear to be multiply-imaged under image resolution
of $0.1'' - 0.8''$.
Note that in our $i$-dropout object search, we do not exclude objects
classified as extended sources. In fact, a few objects in our search
was indeed classified as ``galaxy'' in SDSS processing, 
apparently due to low S/N in
the $z$ band images. 
Therefore, our survey is not biased against lensed quasars.

Recently, \cite{Shioya} report the detection of a faint galaxy
$1.9''$ away from the $z=5.74$ quasar SDSS J1044--0125.
Based on the presence of this galaxy and the non-detection of
a second image under the current resolution, they conclude 
that SDSS J1044--0125 could be magnified by a factor of two,
with a second image too faint to detect. This observation is
consistent with the non-detection of the secondary image in our Keck
imaging.

\subsection{Constraint on the the Slope of Quasar Luminosity Function
from the Magnification Bias}
Wyithe \& Loeb (2002a) and Comerford et al. (2002) calculate the
lensing probability of SDSS $z\sim 6$ quasars. They assume that the quasar
luminosity function is a double power law:
\begin{equation}
\Psi(L) = \frac{\Psi^*/L^*}{(L/L^*)^{\beta_l} + (L/L^*)^{\beta_h}},
\end{equation}
where $L^*$ is the characteristic luminosity, and $\beta_l$ and $\beta_h$ 
are the faint-end and bright-end slope of the quasar luminosity function,
respectively.
Assuming $\beta_l = -1.64$ and $\beta_h = -3.43$, Wyithe \& Loeb (2002)
found that the probability that a quasar in the flux-limited sample of Paper 
I is lensed could be as high as 30\%.
These papers discussed the dependence
of the lensing probability on the assumed slopes, characteristic luminosity,
and minimum image separation.
Comerford et al. (2002) emphasized that the lensing probability
is a strong function of this slope, and that it can approach 100\% for
a slope $\beta_l >4.5$. 
\cite{WL2002b} found a similar dependence on the quasar luminosity function.
Either HST imaging or
high-resolution ground-based imaging of a sample of luminous high-redshift
quasars can be used to place a constraint on this slope.

We carry out the calculations described in Comerford et al. (2002)
to derive the probability of detecting a multiply-imaged quasar in the
high-resolution imaging for each of the SDSS $z\sim 6$ quasars.
We assume that in order to detect a second image from lensing, the
minimum image separation is 0.1, 0.5 and 0.8 for the HST, Keck and
ARC observations, respectively.
The minimum ratio of the primary and second images is assumed to be
the ratio of the quasar and the 10-$\sigma$ detection limit of the image.
Figure 9 shows a contour plot of the probability that {\em no lensed
quasar is detected in high-resolution imaging} as a function of the bright-end
slope of quasar luminosity function $\beta_h$ and the characteristic
absolute magnitude of the luminosity function $M^*_{1450}$ ($\Lambda$-model).
The lines show the 0.32, 0.05, 0.01 and 0.001 probability contours.
As discussed in Comerford et al. (2002) and in
Wyithe \& Loeb (2002b), the probability depends weakly on
the faint-end slope. 
From Figure 9, the fact that no quasar was detected as lensed
in the high-resolution images places a 2-$\sigma$ limit 
on the bright-end slope of the quasar luminosity at $z\sim 6$ at
$\beta = -3.5 \sim -4.5$, depending on the characteristic luminosity
assumed. 
This constraint is general agreement with the result from
directly fitting the luminosity function (\S 4.2) where 
we found $\beta > -3.5$ at 2-$\sigma$
level.
This calculation depends on a number of assumptions made in
the lensing model, such as the population and density profiles of
the lensing galaxy, although these dependences are weak. 
The constraints based on the luminosity distribution of
our sample and on the absence of lensed quasars are completely
independent. They both indicate that the bright-end slope of the quasar 
luminosity function is
not very steep and are consistent with the shallow slope found in Fan et al. (2001c)
for $z\sim 4$ quasars. 

The observation of Shioya et al. (2002) and the possibility of modest
magnification of SDSS J1044--0125 can be used to put an additional constraint
on the lensing probability and the shape of the quasar luminosity function.
We calculate the expected {\em a-posteriori} probability of
SDSS J1044--0125 being magnified by a factor of two or more, regardless of
splitting angles, for different luminosity function shapes.
For $\beta = -4.5$, this probability is 0.74, and it drops to 0.17 and
0.025 for $\beta = -3.5$ and $\beta = -2.5$, respectively.
These probabilities are consistent with the limit on $\beta$ we derived above. 
They show that even without detectable second image under high image resolution,
the possibility that the luminous $z\sim 6$  are modestly magnified
by foreground galaxies is still quite high.
Note that our calculations include all splitting angles between
the two images and all brightnesses of the secondary image. 
By combining deep imaging of quasar environment such as Shioya et al. with HST
high resolution imaging, we could put a stronger constraint on the shape
of the quasar luminosity function. The complete modeling is beyond
the scope of the current paper and will be addressed in detail in the
analysis of our full HST sample. 

Assuming that these luminous quasars are radiating at Eddington limit and
they are magnified by gravitational lensing, if the relation between 
black hole mass and bulge mass observed at $z\sim 0$ (Magorrian et al. 1998,
Gebhardt et al. 2000, Ferrarese \& Merritt 2000),  
these quasars are likely to reside in dark matter halos with $\sim 10^{13} \rm M_{\odot}$,
represent $\sim 5$-$\sigma$ peaks in the density field
(Paper I) and are located on the tail of the mass function at early epochs. 
The comparison between the slope of the quasar luminosity function
and mass function at $z\sim 6$ will put strong constraints on
models of quasar formation at early epochs.
Models associating quasars with dark matter halos, such as the ones by
Haiman \& Loeb (1998),  predict a steep halo mass function
for these quasars ($\beta$ between --5 and --6 for the mass function). 
\cite{WL2002b} developed a semi-analytic model of quasar evolution
using new observations of the relationship between black hole mass
and host galaxy velocity dispersions;
they were able to reproduce the shallow luminosity function slope
at $z>4$, but the predicted slope at $z<2$ is shallower than that
observed in low-redshift quasars.  

The shape of the quasar luminosity function is also crucial in
understanding the contribution of the quasar population to the 
UV ionizing background at high redshift.
In Paper I, we calculated the total emissivity per unit volume of
ionizing photons from the quasar population and compared it with
what is required to keep the universe ionized. 
Figure 10 of Paper I indicates that with $\beta_l > -3.5$ and 
a range of assumptions about the characteristic luminosity and
faint-end slope, quasars
are not likely to produce enough photons to ionize the universe
or to keep it ionized at $z\sim 6$, unless the luminosity function
rises sharply at lower luminosities.

The current SDSS survey can only detect and select
the most luminous $z\sim 6$ quasars with $M_{1450} \lesssim -26.5$.
A deep quasar survey at $z\sim 6$ is needed to understand the
complete picture of high-redshift quasar evolution and its relation
to the reionization history.
The SDSS Southern Survey covers a 200 deg$^2$ region  about 1.5 -- 2  mag
deeper than the SDSS main sample.
Assuming a quasar luminosity function of $\Psi \propto L^{-3}$,
and the density of $z>5.7$ quasars for $z^* < 20$ is 6 objects per
3000 deg$^2$, we expect to find 16 quasars at $z>5.7$ and
$z^* < 22$ in this 200 deg$^2$.
By comparison, we expect to find $\sim 20$ quasars at $z>5.7$ and
$z^* < 20$ in the 10,000 deg$^2$ of main SDSS survey area.

Funding for the creation and distribution of the SDSS Archive has been provided by the Alfred P. Sloan Foundation, the Participating Institutions, the National Aeronautics and Space Administration, the National Science Foundation, the U.S. Department of Energy, the Japanese Monbukagakusho, and the Max Planck Society. The SDSS Web site is http://www.sdss.org/.
The SDSS is managed by the Astrophysical Research Consortium (ARC) for the Participating Institutions. The Participating Institutions are The University of Chicago, Fermilab, the Institute for Advanced Study, the Japan Participation Group, The Johns Hopkins University, Los Alamos National Laboratory, the Max-Planck-Institute for Astronomy (MPIA), the Max-Planck-Institute for Astrophysics (MPA), New Mexico State University, University of Pittsburgh, Princeton University, the United States Naval Observatory, and the University of Washington.
We thank the staff at APO, HET, Calar Alto and Keck for their
expert help.
We acknowledge support from NSF grant PHY 00-70928, a Frank and Peggy
Taplin Fellowship and the University of Arizona (X.F.), NSF grant
AST 00-71091 (M.A.S.) and NSF grant AST 99-00703 (D. P. S.).

\newpage

\begin{deluxetable}{rrr}
\tablenum{1}
\tablecolumns{3}
\tablecaption{Summary of Follow-up Results in the Complete Sample}
\tablehead
{
  & number of objects & percentage
}
\startdata
$z>6$ quasars & 2 & 2.0\% \\
T dwarfs & 9 & 9.2\% \\
M/L dwarfs & 67 & 68.4\% \\
false detections & 20 & 20.4\% \\ \hline
TOTAL & 98 & 100\%
\enddata
\end{deluxetable}

\begin{deluxetable}{cccccc}
\tablenum{2}
\tablecolumns{6}
\tablecaption{Photometric Properties of Three New $z>5.7$ Quasars}
\tablehead
{
object & redshift & $i$ & $z$ & $J$ & SDSS run
}
\startdata
J104845.05+463718.3 & 6.23 $\pm$ 0.05 & 22.38 $\pm$ 0.19 & 19.86 $\pm$ 0.09 & 18.40 $\pm$ 0.05 & 2964  \\
J114816.64+525150.2 & 6.43 $\pm$ 0.05 & 23.30 $\pm$ 0.30 & 20.01 $\pm$ 0.09 & 18.25 $\pm$ 0.05 & 2883  \\
                    &  & 23.86 $\pm$ 0.77 & 20.12 $\pm$ 0.09 & 18.25 $\pm$ 0.05 & 2830 \\
J163033.90+401209.6 & 6.05 $\pm$ 0.03 & 23.38 $\pm$ 0.35 & 20.42 $\pm$ 0.12 & 19.38 $\pm$ 0.10 &  2328  
\enddata
\tablenotetext{}{The SDSS photometry ($i,z$) is
reported in terms of {\em asinh magnitudes} on the AB system.
The asinh magnitude system is defined by Lupton, Gunn \& Szalay (1999);
it becomes a linear scale in flux when the absolute value of the
signal-to-noise ratio is less than about 5. In this
system, zero flux corresponds to 24.4 and 22.8,
in $i$, and $z$, respectively; larger magnitudes refer to negative flux values.
The $J$ magnitude is on a Vega-based system.}
\end{deluxetable}

\begin{deluxetable}{ccccccc}
\tablenum{3}
\tablecolumns{7}
\tablecaption{Continuum Properties of new $z>6$ Quasars}
\tablehead
{
object & redshift & $AB_{1280}$ & $AB_{1450}$ & $M_{1450}$ & $M_{1450}$ & $E(B-V)$ \\
  &  &   &  & ($\Omega$-model) & ($\Lambda$-model) &  (Galactic)
}
\startdata
J104845.05+463718.3 & 6.23 & 19.32 & 19.25 & --27.28 & --27.55  & 0.018\\
J114816.64+525150.2 & 6.43 & 19.10 & 19.03 & --27.55 & --27.82  & 0.023\\
J163033.90+401209.6 & 6.05 & 20.71 & 20.64 & --25.85 & --26.11  & 0.011
\enddata
\end{deluxetable}
\newpage

\begin{figure}[hdt]

\epsfysize=300pt \epsfbox{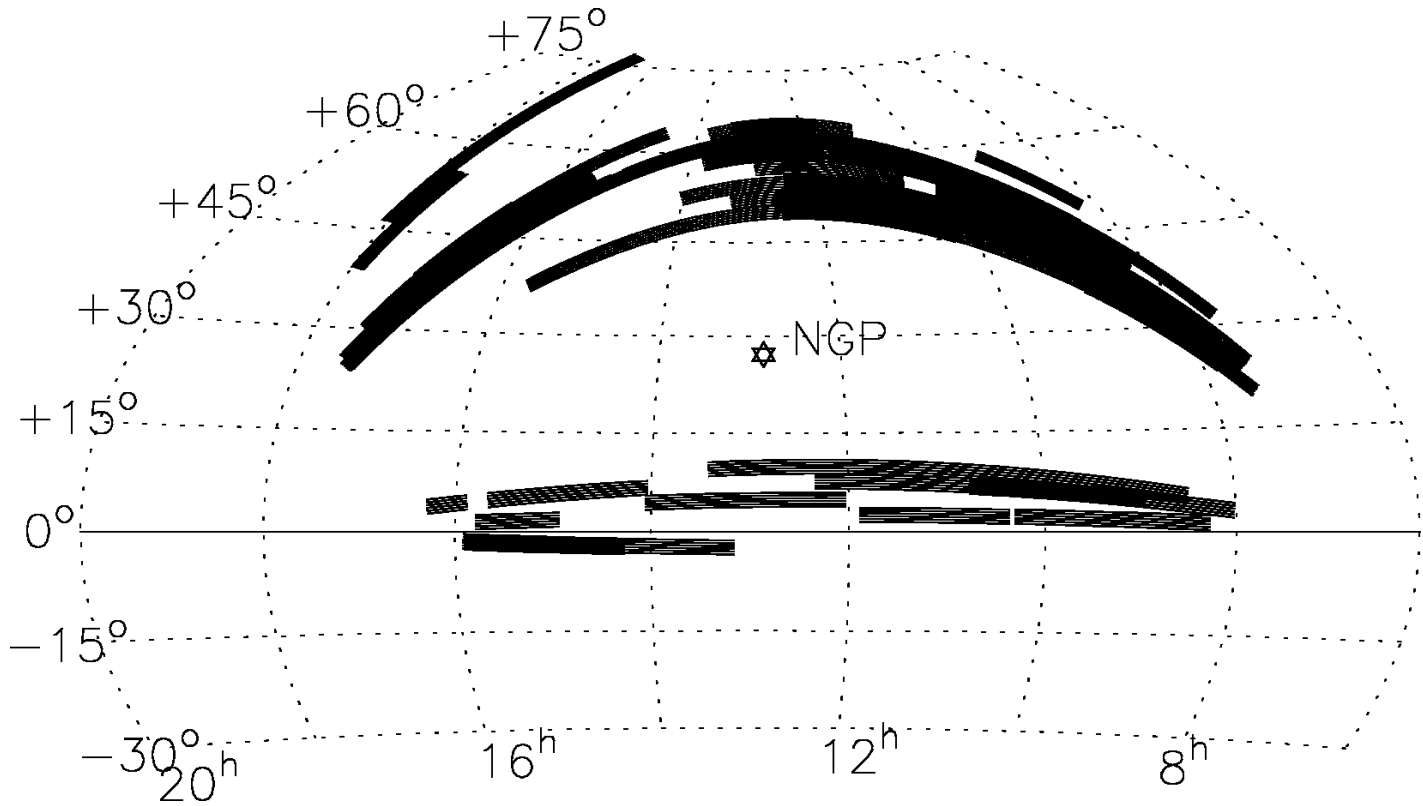}

\caption{Projection of the new area surveyed for $i$-dropout quasars 
in this paper,
in J2000.0 equatorial coordinates. Note that for some SDSS stripes
only one of the two strips is covered by the current survey,
these appear lighter in this figure.
The stripes are each 2.5 deg wide. The total new area surveyed is
1320 deg$^2$.}
\end{figure}

\begin{figure}[hbt]
\epsscale{0.60}
\plotone{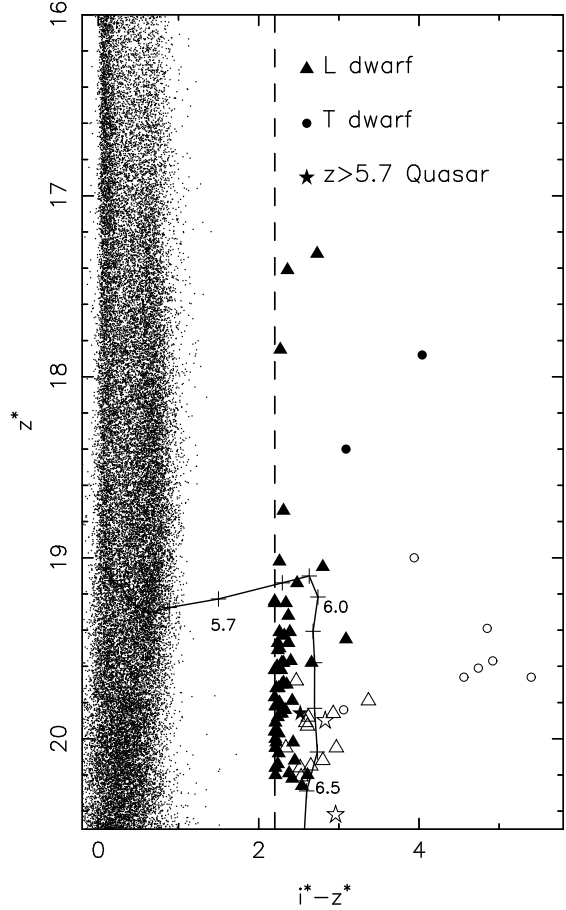}
\caption{The $i^*-z^*$ vs $z^*$ color-magnitude diagram for the $i$-dropout sample. Colors and magnitudes are plotted as
asinh magnitudes measured by SDSS imaging.
The symbols represent the classification
in Table 1: circles: T dwarfs; triangles: L dwarfs;
stars: $z\sim 6$ quasars. Filled symbols are objects
with S/N in the $i$ band higher than 4; open symbols are objects not
detected in the $i$ band at the 4-$\sigma$ level.
The median track of simulated $i^*-z^*$ color and $z^*$ magnitude for
quasars with $M_{1450} = -27$ is also shown as a function of redshift,
with plus signs every 0.1 in redshift.
For comparison, the data for a random sample of 50,000 high-latitude
stars are also shown as dots. The dashed line shows the cut
$i^* - z^* > 2.2$ that we use to select high-redshift quasar
candidates.}
\end{figure}

\begin{figure}[hbt]
\epsscale{0.75}
\plotone{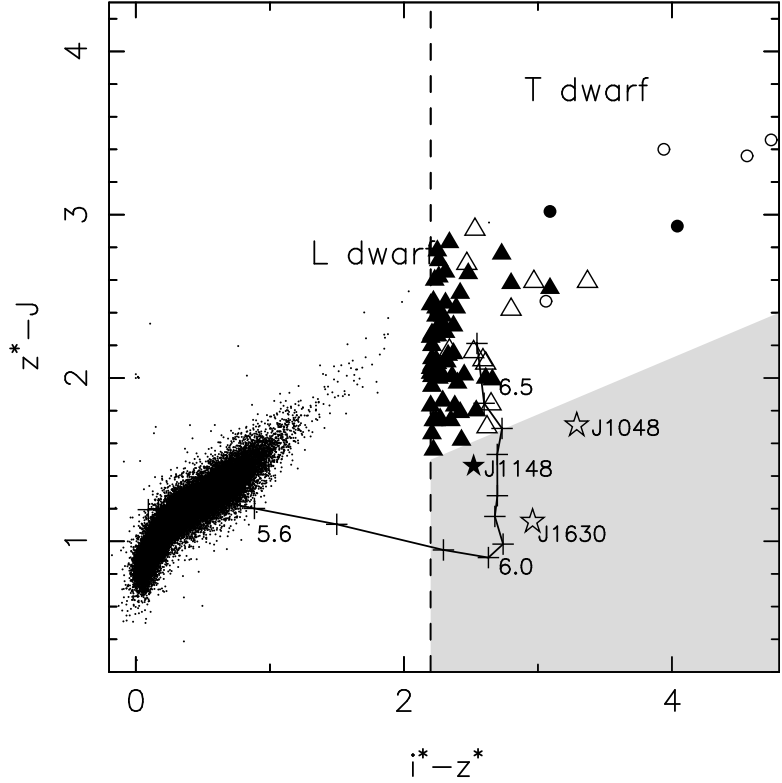}
\caption{$i^*-z^*$ vs. $z^*-J$ color-color diagram for the
$i$-dropout sample. The symbols are the same as in Figure 1.
The median track of simulated quasar colors is shown as a function
of redshift.
The survey selection criteria are illustrated by the shaded area.
Note that we have expanded the color cuts to look for quasars at
higher redshift (c.f. Figure 2 of Paper I).
For comparison, colors of SDSS-2MASS stars in a 50  deg$^2$
area at high latitude are also shown.
}
\end{figure}
\newpage

\begin{figure}[hbt]
\epsfysize=400pt \epsfbox{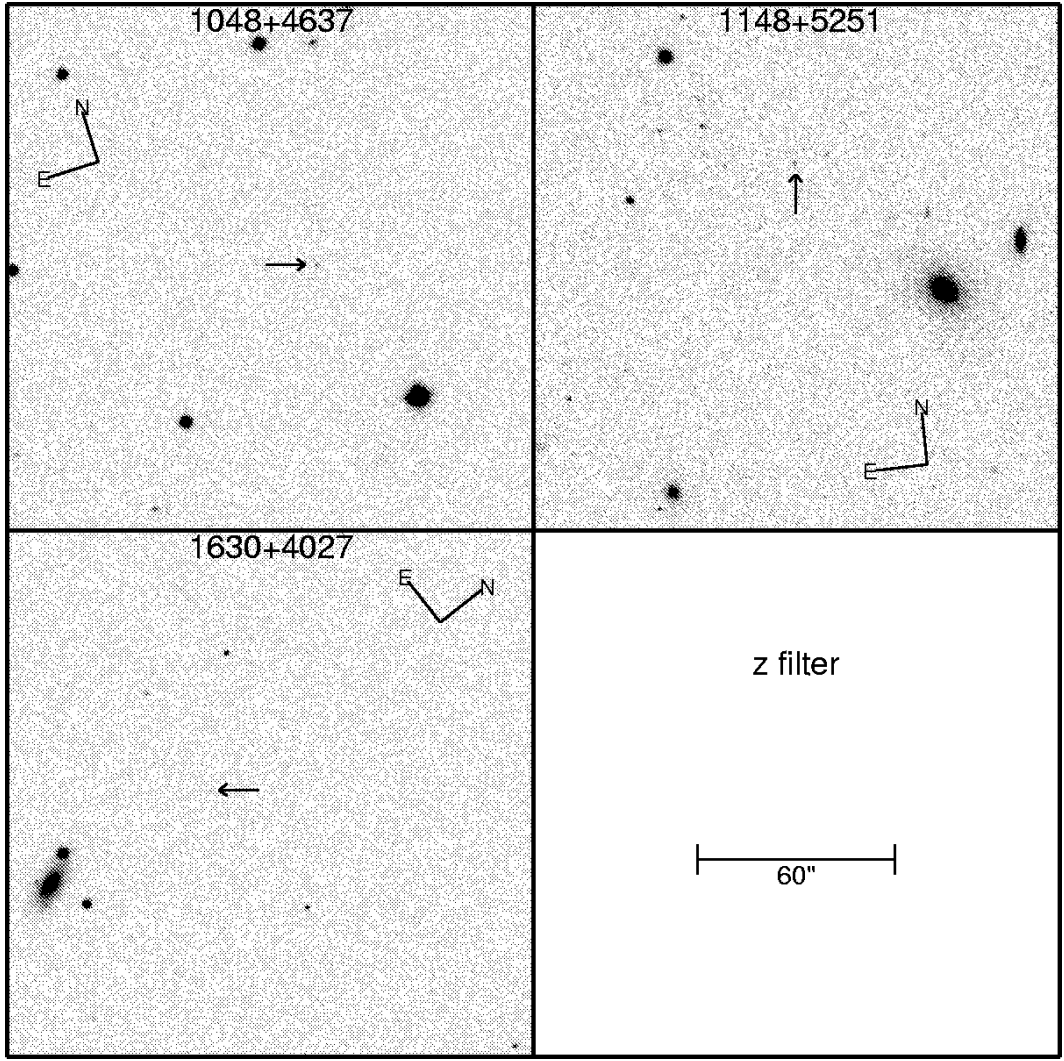}

\vspace{2cm}
\caption{The $z$ finding charts of the three new $z\sim 6$ quasars.
The SDSS $z$ images are shown. The side of the finding chart is
160$''$. The orientation of the finding chart is indicated by 
the North/East direction. }
\end{figure}

\begin{figure}[hbt]
\epsscale{0.70}
\plotone{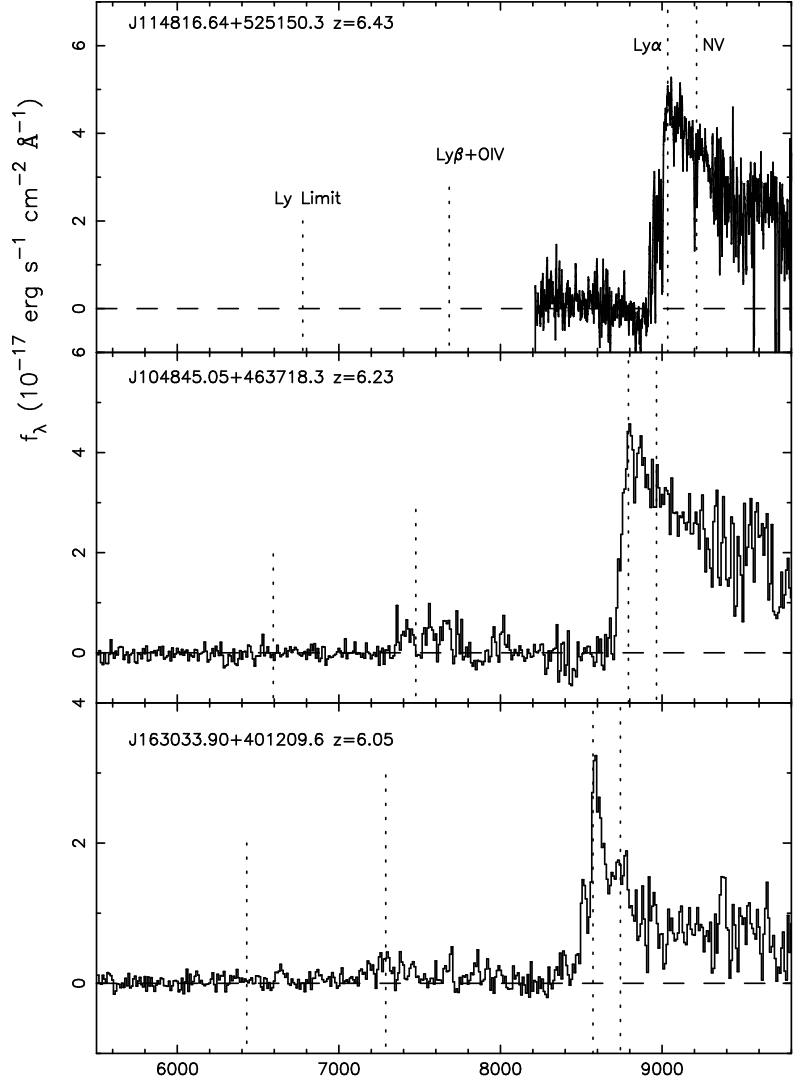}
\caption{Spectra of the three new quasars at $z>6$.
For SDSS J1148+5251, a spectrum taken at the Calar Alto 3.5m is shown, and
binned to 1\AA/pixel;
for SDSS J1048+4637 and SDSS J1630+4012, spectra taken at the HET
are shown, and binned to 5\AA/pixel.
The spectra are flux-calibrated to match the SDSS $z$ photometry.
The original discovery spectra of these three objects were taken
with the ARC 3.5m.
}
\end{figure}

\begin{figure}[hbt]
\vspace{-3cm}

\epsscale{1.00}
\plotone{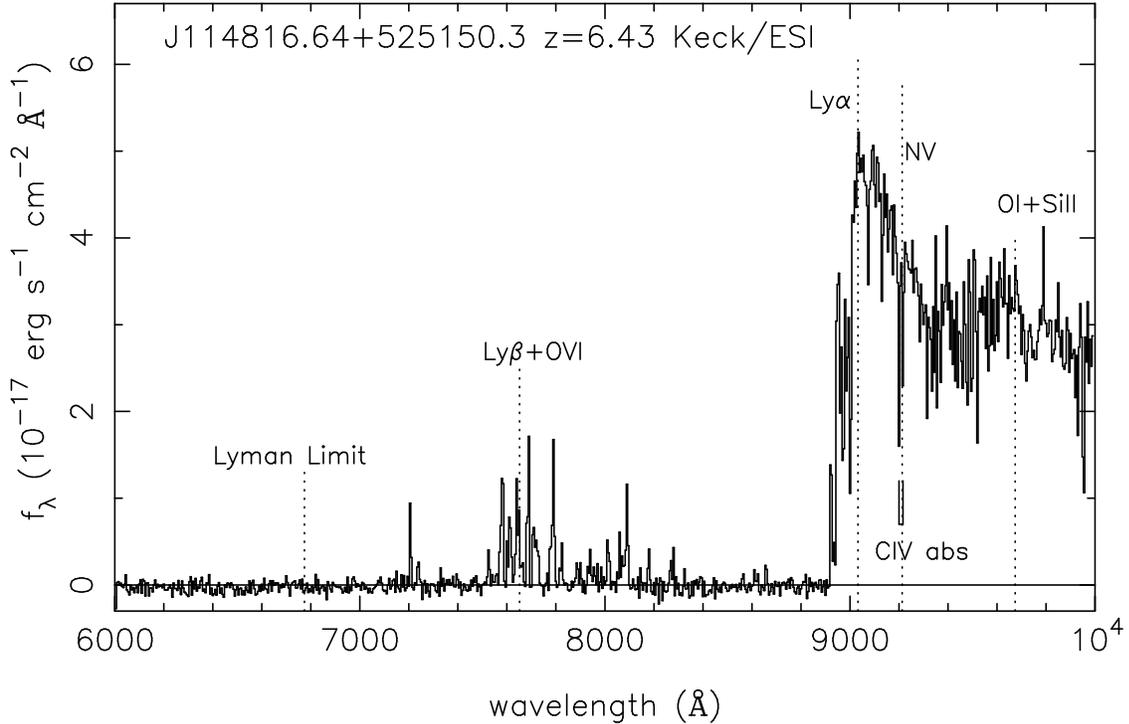}
\caption{A Keck/ESI spectrum of J1148+5251. It is a 3 hour exposure under
very marginal conditions (extinction $>$ 1 magnitude). The spectrum
is binned to 2\AA/pixel, and is flux-calibrated to match the
SDSS $z$ photometry.
Note a strong CIV doublet absorber at $\sim 9200$\AA\ ($z=4.95$).
Also note a complete Gunn-Peterson trough on the blue side of the
Ly$\alpha$ emission where no flux is detected (see Fan et al. 2003).}
\end{figure}

\begin{figure}[hbt]
\vspace{-2cm}

\epsscale{1.00}
\plotone{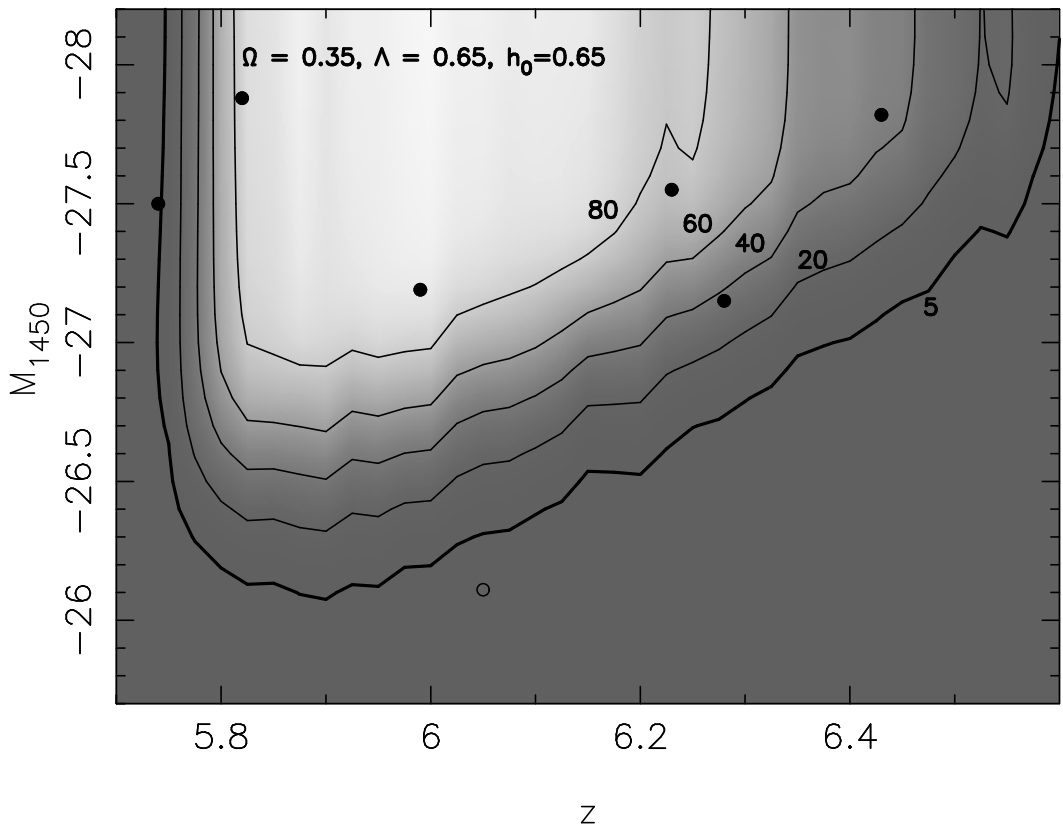}
\caption{
The selection probability of $i$-dropout quasars
as a function of redshift and
luminosity, for the $\Lambda$ models.
Probability contours of 5\%, 20\%, 40\%, 60\% and 80\% are shown.
The large dots represent the locations of the six quasars in the complete 
sample.  SDSS J1630+4012 is shown as an open circle as it is fainter than
the magnitude limit of the complete sample.
}
\end{figure}

\begin{figure}[hbt]
\epsscale{0.80}
\plotone{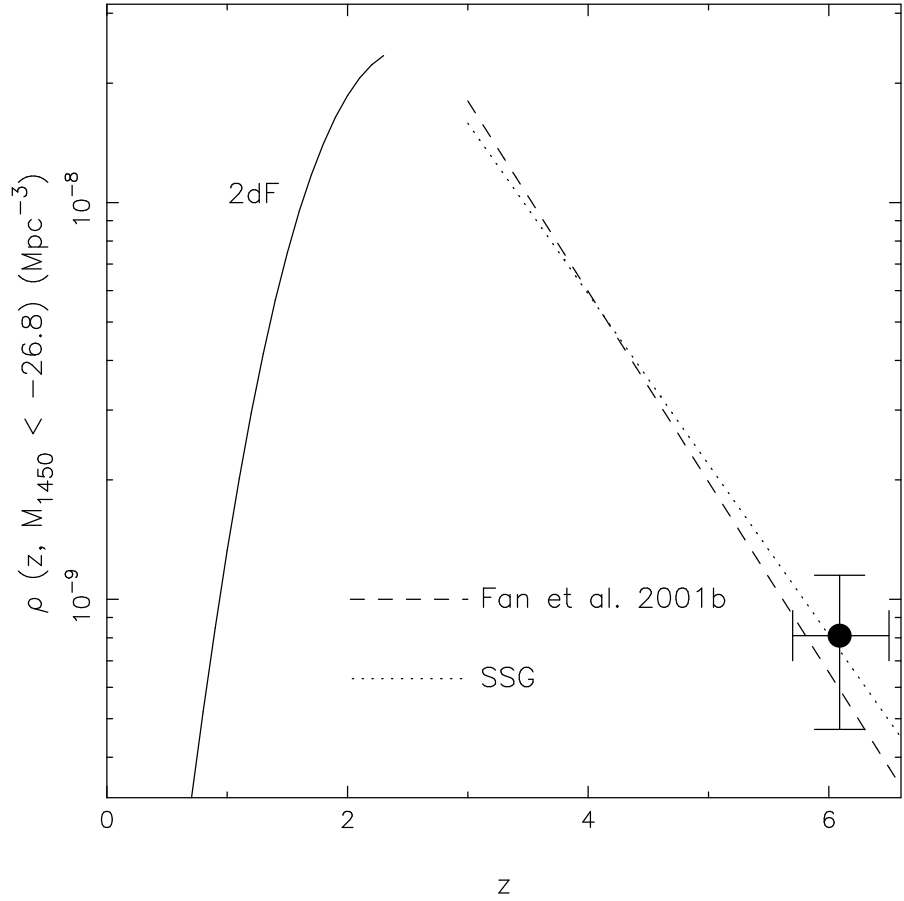}
\caption{
The evolution of quasar comoving spatial density at
$M_{1450} < -26.8$ in the $\Omega=1$ model.
The large dot represents the result from this survey.
The dashed and dotted lines are the best-fit models
from Fan et al.~(2001b) and Schmidt et al.~(1995, SSG),
respectively.
The solid line is the best-fit model from the 2dF survey
(Boyle et al.~2000) at $z<2.5$.
}

\end{figure}

\begin{figure}[hbt]
\vspace{-2cm}

\epsscale{1.00}
\plotone{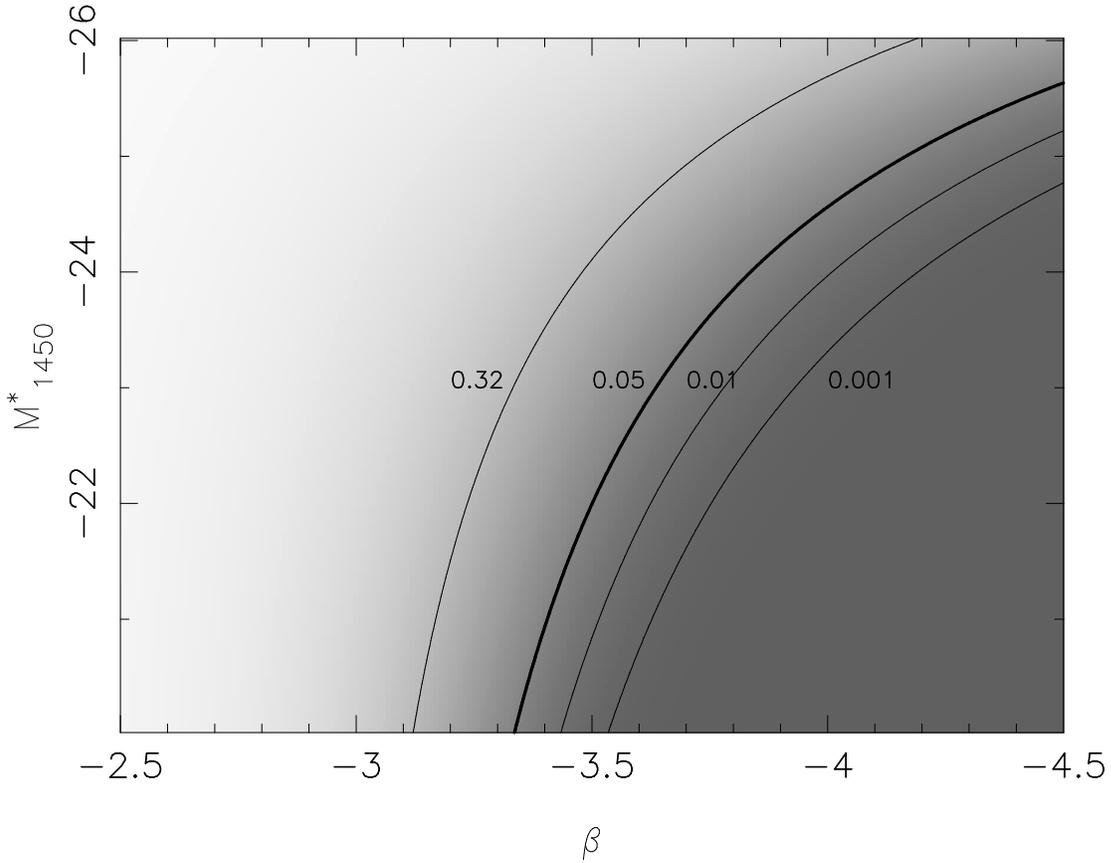}

\caption{Probability that no lensed quasar is detected among the
seven SDSS quasars at $z>5.7$ under the imaging resolutions
described in the paper ($0.1'' - 0.8''$)
It is shown as a function of the bright-end slope $\beta$ and 
characteristic luminosity ($M^*_{1450}$) of the quasar luminosity
function at $z\sim 6$.
The contours show the probabilities of 32\%, 5\%, 1\%, and 0.1\% that
no lensed quasar is detected.
None of the quasars appear to be lensed under these conditions.
Based on this plot, the quasar luminosity function is likely to
be flatter than $L^{-3.5 \sim -4.5}$ with 2-$\sigma$ confidence (95\%),
depending on the characteristic luminosity assumed.}
\end{figure}

\end{document}